\documentclass[journal]{IEEEtran}
\usepackage{CJKutf8}
\usepackage{amsfonts}
\usepackage{amssymb}
\usepackage{url}
\usepackage{array}
\usepackage{multirow,multicol}
\usepackage{threeparttable}
\usepackage{tablefootnote}
\usepackage[super]{nth}
\usepackage{cite}
\usepackage{makecell}
\usepackage{booktabs}
\usepackage{amsmath}
\usepackage{xcolor}
\usepackage{graphicx}
\usepackage{mathrsfs}
\usepackage{tabularx}
\usepackage{verbatim}
\usepackage{bbding}
\usepackage{diagbox}
\usepackage{setspace}
\usepackage{bm} 
\usepackage{algorithmicx,algorithm}
\usepackage[noend]{algpseudocode}

\ifCLASSINFOpdf
\else
\fi
\begin{document}
%
\title{Label-Synchronous Neural Transducer for Adaptable Online E2E Speech Recognition}
%
%

%

\author{Keqi Deng,~\IEEEmembership{Student Member,~IEEE,}
        Philip C. Woodland,~\IEEEmembership{Fellow,~IEEE}
\thanks{Keqi Deng is funded by the Cambridge Trust. This work has been performed using resources provided by the Cambridge Tier-2 system operated by the University of Cambridge Research Computing Service (www.hpc.cam.ac.uk) funded by EPSRC Tier-2 capital grant EP/T022159/1.}
\thanks{Keqi Deng and Philip C. Woodland are with the Department of Engineering, University of Cambridge, CB2 1TN Cambridge, U.K. (e-mail: kd502@cam.ac.uk; pw117@cam.ac.uk).}}
\maketitle

\begin{abstract}
Although end-to-end (E2E) automatic speech recognition (ASR) has shown state-of-the-art recognition accuracy, it tends to be implicitly biased towards the training data distribution which can degrade generalisation.
This paper proposes a label-synchronous neural transducer (LS-Transducer), which provides a natural approach to domain adaptation based on text-only data.
The LS-Transducer extracts a label-level encoder representation before combining it with the prediction network output. 
Since blank tokens are no longer needed,
the prediction network performs as a standard language model, which
can be easily adapted using text-only data.
An Auto-regressive Integrate-and-Fire (AIF) mechanism is proposed to generate the label-level encoder representation while retaining low latency operation that can be used for streaming. In addition, a streaming joint decoding method is designed to improve ASR accuracy while retaining synchronisation with AIF. Experiments show that compared to standard neural transducers, the proposed LS-Transducer gave a 12.9\% relative WER reduction (WERR) for intra-domain LibriSpeech data, as well as 21.4\% and 24.6\% relative WERRs on cross-domain TED-LIUM 2 and AESRC2020 data with an adapted prediction network.

\end{abstract}

\section{Introduction}
\label{sec:intro}
\IEEEPARstart{T}{he}
hybrid deep neural network and hidden Markov model (DNN-HMM) \cite{5740583, Hinton2012DeepNN, Vesel2013SequencediscriminativeTO} approach for automatic speech recognition (ASR)
is a widely-used deep learning-based framework, which contains several separately optimised modules, including the acoustic model, pronunciation lexicon, context dependency model \cite{Young1994TreeBasedST}, and language model (LM). 
However, the separately optimised models make it hard to achieve a globally optimised system \cite{Wang2019AnOO}. End-to-end (E2E) ASR simplifies the modelling pipeline and integrates 
the separate modules used by the DNN-HMM approach \cite{6638947}.
Notable E2E ASR methods include connectionist temporal classification (CTC) \cite{graves2006connectionist}, the neural transducer \cite{Graves2012SequenceTW}, and the attention-based encoder–decoder (AED) \cite{Vaswani2017}.
Among these techniques, the neural transducer provides a natural approach for streaming ASR that can give a high accuracy, and has become popular for industrial applications \cite{SIP-2021-0050}.

The hybrid DNN-HMM framework is, however, still used in many industrial ASR systems \cite{SIP-2021-0050}. Several practical advantages of hybrid systems contribute to this, including low-latency streaming and domain adaptation capabilities \cite{SIP-2021-0050}. In the past few years, the streaming features of E2E ASR have been extensively explored \cite{9413535} and neural transducers can replace hybrid DNN-HMMs in some cases \cite{li20_interspeech}. However, the hybrid DNN-HMM approach
is highly modular and contains an explicit independent LM, making it straightforward to bias the recognition system to unseen domains using text-only data. The neural transducer, which is the most commonly deployed E2E model, is still 
weaker than HMM-based systems in this regard. This is because E2E ASR models jointly learn acoustic and linguistic information \cite{LI202312} and do not have an explicit LM that can be used for flexible domain adaptation with text-only data. 
For the standard neural transducer, in which speech is decoded on a per-frame basis i.e. frame-synchronously, blank tokens are used to augment the output sequence thus allowing the frame-level encoder output to be combined with the label-level prediction network output \cite{Graves2012SequenceTW}. However, blank token generation means that the prediction network cannot be viewed as an explicit LM \cite{9054419} due to the inconsistency with the LM task \cite{9054419, Chen2021FactorizedNT} and thus poses a challenge to
text-only domain adaptation.

The motivation of this paper is to modify the standard neural transducer by 
enabling an acoustic encoder representation to be directly combined with the prediction network output at the label level and hence not need blank tokens. Therefore, operation is label-synchronous and
the prediction network performs as a standard LM. This makes it straightforward to biased the model to previously unseen domains using text-only data. 
In this paper, a label-synchronous neural transducer (LS-Transducer) is proposed\footnote{An earlier and less complete description of the work is available at \cite{deng2023labelsynchronous}.} that provides a natural approach to domain adaptation, while retaining the valuable streaming property and E2E training simplicity.

The main contributions of this work can be summarised in four key parts:
\begin{itemize}
    \item {\it LS-Transducer:} A label-synchronous neural transducer (LS-Transducer) is proposed that extracts a label-level encoder representation before combining it with the prediction network output. The LS-Transducer improves E2E ASR domain adaptation while maintaining low ASR error rates and good streaming properties.
    \item {\it Auto-regressive Integrate-and-Fire (AIF):} In order to extract the label-level encoder representation for the LS-Transducer, an AIF mechanism is proposed, which is extended from the Continuous Integrate-and-Fire (CIF) \cite{9054250} approach but with improved efficiency and robustness to inaccurate unit boundaries. 
    \item {\it Prediction Network Adaptation:} The LS-Transducer prediction network performs as an explicit LM, so can be easily fine-tuned on target-domain text.
    \item {\it Streaming Joint Decoding:} This paper also proposes a streaming joint decoding method to enhance the accuracy of the LS-Transducer by utilising an online CTC prefix score, which is  synchronised with the AIF alignment through a simple but effective modification to the standard CTC prefix score.
\end{itemize}

Experiments with ASR models trained 
on LibriSpeech data \cite{7178964} show that the proposed LS-Transducer gives reduced WERs over standard neural transducer models for both intra-domain and cross-domain scenarios. 

The rest of this paper is organised as follows. Section~\ref{related work} reviews related works including CIF on which the AIF technique is based. Section~\ref{LST_method} describes the proposed LS-Transducer framework. 
Section~\ref{setup} details the experimental setups and Section~\ref{results}
presents the results. Finally,
Section~\ref{conclusions} concludes.

\section{Related Work}
\label{related work}
\subsection{Neural Transducer Models}
\label{nt}
The neural transducer \cite{Graves2012SequenceTW} has gained widespread interest and is the leading E2E model deployed in industry \cite{SIP-2021-0050}. The neural transducer removes the independence assumption in CTC between output tokens by conditioning on previous non-blank output tokens \cite{Higuchi2022BERTMC}. 
When aligning input speech and output token sequences,
the neural transducer aligns the sequences at the frame level by inserting a blank token to augment output sequences. Consequently, the neural transducer output is conditioned on the speech sequence up to the current time step, providing a natural approach for streaming ASR \cite{SIP-2021-0050}. This is in contrast to the AED which relies on a global attention mechanism that hampers streaming processing or incurs significant latency \cite{wang20v_interspeech},

The neural transducer contains an encoder network, a prediction network, and a joint network. The encoder network extracts an acoustic representation $\bm{h}_t^{\rm enc}$ from input speech $\bm{x}$.
The encoder network can use a long short-term memory (LSTM)\cite{Hochreiter1997Long}, Transformer \cite{Vaswani2017}, or Conformer \cite{gulati20_interspeech} structure. However, when aimed at streaming, strategies like the chunk-based or lookahead-based method \cite{li20_interspeech} need to be employed to achieve a streaming Transformer/Conformer encoder.

The prediction network allows the neural transducer to capture causal dependencies in the output by generating a representation $\bm{h}_n^{\rm pre}$ from previous non-blank tokens $y_{1:n-1}$. The prediction network is an auto-regressive structure and can employ an RNN \cite{Graves2012SequenceTW}, unidirectional Transformer \cite{Zhang2020TransformerTA} or even only an embedding layer \cite{9054419}.
The prediction network normally has a similar structure to an LM, but it does not perform as an explicit LM because it also needs to predict blank tokens, which is inconsistent with the LM task \cite{Chen2021FactorizedNT}.

The joint network combines $\bm{h}_t^{\rm enc}$ and $\bm{h}_n^{\rm pre}$ at the frame level
with fully-connected (FC) networks and the output logits $\bm{l}_{t,n}$ can be computed as:
\begin{equation}
    \bm{l}_{t,n} = {\rm FC}(\Psi({\rm FC}(\bm{h}_t^{\rm enc})+{\rm FC}(\bm{h}_n^{\rm pre})))
\end{equation}
where $\Psi$ is a non-linear activation function and the predicted probability of the $k$-th token is obtained by applying a softmax function to the logits $\bm{l}_{t,n}$:
\begin{equation}
    p(y_n=k|\bm{x}_{1:t}, {y}_{1:n-1})={\rm softmax}(\bm{l}_{t,n})
\end{equation}
where $\bm{x}_{1:t}$ denotes the speech sequence up to frame $t$.
The neural transducer loss function $\mathcal{L}_{\rm nt}$ is defined as the negative log-likelihood of the target text sequence $\bm{y}$ of length $N$:
\begin{eqnarray}
&p(\bm{a}|\bm{x})\approx\prod_{u=1}^{T+N}p(a_u|{\rm A}(a_{1:u-1}),\bm{x})\\
  &\mathcal{L}_{\rm nt}=-{\rm ln}\sum_{\bm{a} \in A^{-1}}p(\bm{a}|\bm{x})
\end{eqnarray}
where $T$ is the total length of $\bm{x}$ and
${\rm A}$ is a  collapsing function that maps all alignment paths $\bm{a}$ to the target text sequence.

\subsection{Continuous Integrate-and-Fire (CIF)}
\label{sec:cif}
\begin{figure}[b]
    \centering
    \vspace{-0.5cm}
    \includegraphics[width=86mm]{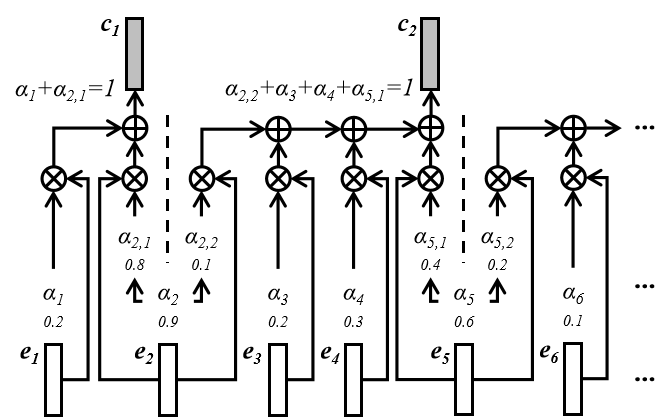}
    \vspace{-0.1cm}
    \caption{Example of CIF \cite{9054250}. $\bm{\oplus}$ and $\bm{\otimes}$ denote addition and multiplication. $\mathbf{E}\text{=}(\bm{e_1}, \cdots, \bm{e_T})$ denotes encoder output and $\bm{\alpha}\text{=}(\alpha_1, \cdots, \alpha_T)$ represents predicted weights whose example values are $(0.2, 0.9, 0.2, 0.3, 0.6, 0.1 \cdots)$ .
    }
    \vspace{-0.3cm}
    \label{cif}
\end{figure}
CIF \cite{9054250} is presented as background since the AIF mechanism in the LS-Transducer is an extension of CIF. 
CIF estimates a monotonic alignment for streaming ASR. In CIF, the first step involves learning a weight $\alpha_t$ for each encoder output frame $\bm{e}_t$. To obtain this weight,
a one-dimensional scalar is generated from each encoder output $\bm{e}_t$ through convolutional or fully-connected layers \cite{9054250} or even directly using a particular element of $\bm{e}_t$ \cite{9398531}.  Using a sigmoid function, the weight $\alpha_t$ is then computed from this scalar.
Next, the CIF mechanism accumulates the weights over time and integrates the acoustic representation via a weighted sum. The accumulation continues until the accumulated weight is above a threshold of 1.0. At this point,
the current weight $\alpha_t$ is split into two parts: one part is used to make the current accumulated weight be exactly $1.0$, while the remainder is used for the integration of the next label. CIF then ``fires" the integrated acoustic representation $\bm{c_j}$ corresponding to label $y_j$ and resets the accumulation.

As shown the example in Fig.~\ref{cif}, if the weights $(\alpha_1, \cdots, \alpha_T)$ generated by CIF are $(0.2, 0.9, 0.2, 0.3, 0.6, 0.1 \cdots)$, the $\alpha_2=0.9$ needs to be divided into $\alpha_{2,1}=0.8$ and $\alpha_{2,2}=0.1$, so that $\alpha_1+\alpha_{2,1}=1.0$  and $\bm{c}_1=0.2\bm{e}_1+0.8\bm{e}_2$ can be emitted. The same situation will also occur when $\alpha_5=0.6$, which needs to be divided into $\alpha_{5,1}=0.4$ and $\alpha_{5,2}=0.2$, so that $\alpha_{2,2}+\alpha_3+\alpha_4+\alpha_{5,1}=1.0$ and $\bm{c}_2=0.1\bm{e}_2+0.2\bm{e}_3+0.3\bm{e}_4+0.4\bm{e}_5$ can be emitted. The calculation of $\bm{c}_3$, $\bm{c}_4$, etc., are similar and applied until the end of the encoder output.

During training, a scaling strategy is used to ensure that the integrated acoustic representations $\mathbf{C}$=$(\bm{c}_1, \cdots, \bm{c}_L)$ have the same length $L$ as the target sequence. This strategy involves computing a scaled weight, $\hat{\alpha}_t$, which is obtained by $\hat{\alpha}_t$=$\alpha_t \cdot (L/\sum_{i=1}^T\alpha_i)$. By using $\hat{\alpha}_t$ instead of $\alpha_t$ to extract $\mathbf{C}$, the length is effectively controlled. However, during decoding, the length of integrated representations is solely determined by the weight accumulation $\sum_{i=1}^T\alpha_i$. Hence, a quantity loss $\mathcal{L}_{\rm qua}=|\sum_{i=1}^T\alpha_i - L|$, defined as the absolute difference between $\sum_{i=1}^T\alpha_i$ and the target length $L$, is used to supervise CIF to extract a number of integrated representations close to $L$.

Note that CIF doesn't always locate the real acoustic boundaries and accurately estimate the text length \cite{9688157, Yao2022WaBERTAL}, especially when applied to units like BPE \cite{gage1994} in English ASR\footnote{CIF gained popularity in Mandarin ASR tasks because Chinese characters correspond to clear syllable boundaries. However, CIF suffers from performance degradation on English ASR due to difficulty in locating boundaries for BPE units, as mentioned in \cite{9688157}. While preliminary experiments show that the proposed LS-Transducer is very effective on Mandarin data, this paper focuses on English ASR.}. 
In addition, since the scaling strategy is used during training, a mismatch exists between training and decoding. Moreover, CIF is a sequential method \cite{Yao2022WaBERTAL, Deng2022ImprovingCS} since it needs to know the time step that the previous label representation was emitted, before the weight accumulation mechanism is reset for the next label.
This can lead to reduced training efficiency.

\subsection{Text Domain Adaptation}
Various methods have been developed for E2E ASR domain adaptation using text-only data.
One solution involves LM fusion where an external LM is integrated into the E2E ASR system \cite{chorowski2015attention, sriram18_interspeech}, of which the most commonly used is shallow fusion \cite{chorowski2015attention}. However, the E2E ASR model implicitly learns an internal LM that characterises the source domain training data distribution \cite{9415039}, which causes a mismatch when decoded on unseen domains. To solve this issue, this internal LM can be estimated \cite{9415039, 9746480, 9053600, zeineldeen21_interspeech, 9383515, 9746948}. For example, HAT \cite{9053600} was proposed to estimate the internal LM by removing the acoustic encoder effect from the neural transducer. Nonetheless, estimating the internal LM score increases the complexity of decoding and achieving accurate estimation is not always feasible due to domain mismatch \cite{tsunoo22_interspeech}. More recent studies \cite{Chen2021FactorizedNT, meng22_interspeech, Meng2022ModularHA} such as the factorised neural transducer \cite{Chen2021FactorizedNT},
have investigated fine-tuning the internal LM using target-domain text. However, this approach
can lead to intra-domain performance degradation \cite{Chen2021FactorizedNT}. The use of Kullback-Leibler divergence regularisation mitigates this issue but limits how much the internal LM learns the target domain \cite{meng22_interspeech, Meng2022ModularHA}. Another solution involves using Text-to-Speech (TTS) to synthesise speech from the target-domain text, which is then employed for fine-tuning the ASR models \cite{deng21_interspeech, Zheng2020UsingSA}. However, this method 
incurs significant computational cost and lacks flexibility for fast adaptation \cite{Chen2021FactorizedNT}.

\section{Label-synchronous Neural Transducer}
\label{LST_method}

\begin{figure}[th]
    \centering
    \vspace{-0.1cm}
    \includegraphics[width=70mm]{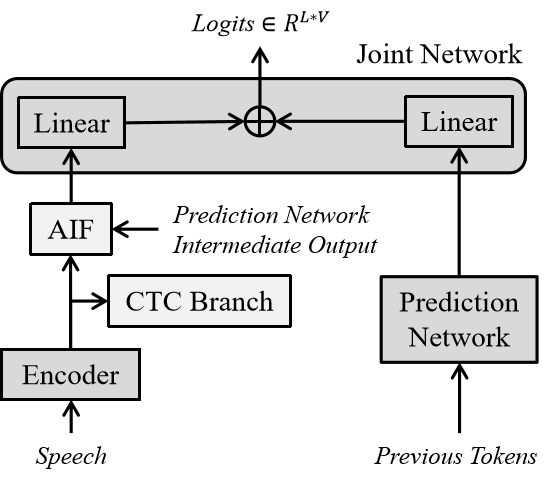}
    \vspace{-0.2cm}
    \caption{Illustration of the proposed LS-Transducer. Linear denotes linear classifier. The output $logits$ is a label-level two-dimensional matrix, where $L$ and $V$ are the length and vocabulary size. $\bm{\oplus}$ denotes addition.}
    \vspace{-0.2cm}
    \label{ls-t}
\end{figure}

As illustrated in Fig.~\ref{ls-t}, the proposed LS-Transducer includes the AIF mechanism to extract a label-level encoder representation before combining it with the prediction network output. This is the main difference with the standard neural transducer which directly combines the frame-level encoder output with the label-level prediction network output. The joint network in the LS-Transducer then adds the logits obtained from the AIF and prediction network outputs through linear fully-connected layers. This design enables the prediction network to be flexibly biased to unseen domains on text-only data, without affecting other parts of the model.
Note that the joint network output, as shown in Fig.~\ref{ls-t},
is a 2-dimensional matrix $\mathbb{R}^{L\cdot V}$, which differs from the standard neural transducer where the output is a 3-dimensional tensor $\mathbb{R}^{T \cdot L\cdot V}$ with an extra time dimension.

During training, with the help of the proposed AIF mechanism, the logits (as in Fig.~\ref{ls-t}) in the LS-Transducer will have the same length as the target sequence, therefore the cross-entropy (CE) loss $\mathcal{L}_{\rm ce}$ can be
computed between them and used as the training objective. Computing this CE loss can also help save a considerable amount of memory compared to the RNN-T loss that is computed based on three-dimensional tensors \cite{SIP-2021-0050}. In addition, the AIF mechanism calculates the quantity loss $\mathcal{L}_{\rm qua}$ to learn an explicit speech-text alignment, which will be described in detail in Section~\ref{aif}. Inspired by \cite{9688251}, which shows that CTC \cite{graves2006connectionist} always helps model training for both AED and neural transducers, CTC-based supervision $\mathcal{L}_{\rm ctc}$ is also used by the encoder in the LS-Transducer. Therefore, the overall training objective $\mathcal{L}_{\rm lst}$ of the LS-Transducer is computed as:
 \begin{equation}
    \mathcal{L}_{\rm lst}=\gamma \mathcal{L}_{\rm ctc}+(1-\gamma) \mathcal{L}_{\rm ce}+\mu \mathcal{L}_{\rm qua}\cdot L \label{obj}
\end{equation}
where $L$ is the target length, and $\gamma$ and $\mu$ are hyper-parameters.

\subsection{Auto-regressive Integrate-and-Fire (AIF)}
\label{aif}
\begin{figure}[t]
    \centering
    \includegraphics[width=88mm]{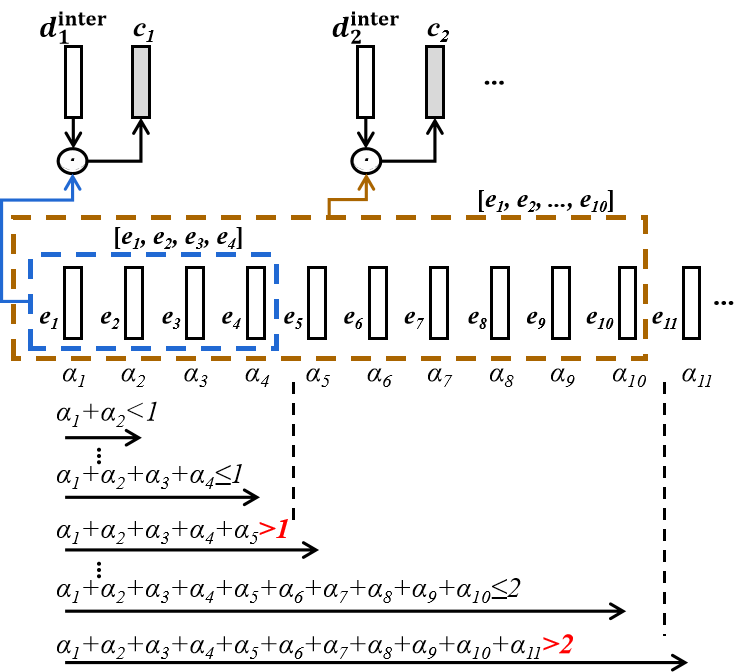}
    \caption{Illustration of the proposed AIF. \textcircled{·} denotes dot-product attention, whose query $\bm{d_j}^{\rm inter}$ is the intermediate output of the prediction network.} 
    \label{fig:aif}
\end{figure}

The AIF mechanism is proposed in this paper, which extracts label-level representations $\mathbf{C}=(\bm{c}_1, \cdots, \bm{c}_L)$ from the acoustic encoder output $\mathbf{E}=(\bm{e}_1, \cdots, \bm{e}_T)$. 
Extended from CIF \cite{9054250}, AIF retains the streaming property and also uses accumulated weights $\alpha_t$ to locate boundaries and thus decide when to fire a label-level representation $\bm{c}_j$. However, AIF resolves a number of issues present in CIF, thus improving the LS-Transducer recognition accuracy. The main difference to CIF is that AIF uses dot-product attention instead of the weights $\alpha_t$ to extract the $\bm{c}_j$, after locating the boundaries based on the accumulated weights $\alpha_t$.

To be more specific, as shown in Fig.~\ref{fig:aif}, AIF first computes a weight $\alpha_t$ for each encoder output frame $\bm{e}_t$, and this weight can be obtained using a sigmoid function, after transforming $\bm{e}_t$ into a scalar by neural networks or even simply selecting a particular element. The next step is locating the boundaries corresponding to the ASR modelling unit. To decide when to fire the label-level representation $\bm{c}_j$ where $j\in (1, L)$,
the weight $\alpha_t$ will be accumulated from left to right until it exceeds $j$ ($j$ is both the index and threshold for $\bm{c}_j$), and then the time step of the located boundary for $\bm{c}_j$ is recorded as $T_j$+1.
If the $j$ isn't reached until all $T$ frames have been read (i.e. $\sum_{i=1}^T\alpha_i\leq j$), $T_j=T$. When firing the label-level representation $\bm{c}_j$, AIF employs dot-product attention, where $\textbf{E}_{1:T_j}$ is used as the keys and values:
 \begin{equation}
    \bm{c_j} = {\rm softmax}(\bm{d}_j^{\rm inter}\cdot{ {\textbf{E}_{1:T_j}}^{\top}})\cdot{\textbf{E}_{1:T_j}} \label{label}
\end{equation}
where the query $\bm{d}_j^{\rm inter}$ is the prediction network intermediate output at the $j$-th step. As shown in Fig.~\ref{fig:aif}, after the $\bm{c}_j$ is extracted, the accumulation of the weight $\alpha_t$ will continue to locate the 
boundaries of $\bm{c}_{j+1}$, this process is carried out incrementally until the last $\bm{c}_L$ is obtained.

As shown in the example in Fig.~\ref{fig:aif}, when generating the first representation $\bm{c}_1$, the accumulated weight $\alpha_t$ exceeds $1$ at the $5$-th time step (i.e. $\sum_{i=1}^5\alpha_i>1$ and $\sum_{i=1}^4\alpha_i\le1$),
thus $\mathbf{E}_{1:4}$ is used as the keys and values to extract $\bm{c}_1$ with $\bm{d}_1^{\rm inter}$ as the query. After that,
the weights $\alpha_t$ continue to be accumulated to find the time step when the accumulation exceeds $2$, which is at the $11$-th time step in this example. Therefore, $\mathbf{E}_{1:10}$ is used as the keys and values to extract $\bm{c_2}$ with query $\bm{d}_2^{\rm inter}$. Subsequent extraction for $\bm{c}_3$, $\bm{c}_4$, etc., follows a similar rule.

During training, in order to encourage AIF to learn accurate speech-text alignments and thus locate the correct boundaries, an explicit objective i.e. quantity loss $\mathcal{L}_{\rm qua}$ will be computed:
\begin{equation}
    \mathcal{L}_{\rm qua}=|\sum_{i=1}^T\alpha_i - L| \label{bpe:qua}
\end{equation}
This alignment learning is explicit and independent of recognition, as these weights $({\alpha_1}, \cdots, {\alpha_T})$ are not used to extract 
label-level representations $\mathbf{C}=(\bm{c}_1, \cdots, \bm{c}_L)$ and thus predict the target labels, hence distinguishing it from other E2E methods such as standard neural transducers, CTC, and AED.

In general, AIF generates the label-level representation in an auto-regressive manner, which has many advantages compared to conventional CIF.
First of all, as mentioned in Sec.~\ref{sec:cif}, the length of label-level representation $\mathbf{C}$ extracted by CIF is solely determined by the value of accumulated weights $\sum_{i=1}^T\alpha_i$ and thus needs to employ a scaling strategy during training. However, this scaling strategy relies on the ground truth of target length and is not accessible during decoding, causing a mismatch between training and decoding. However, in AIF, the length of $\mathbf{C}$ is decided by the number of queries, so the scaling strategy is not used and the mismatch issue does not arise.
Second, AIF has a higher training speed due to its ability to generate label-level representations in parallel with teacher forcing by masking certain attention weights, while, CIF, as discussed in Sec.~\ref{sec:cif}, is a sequential approach.
Third, although the boundaries located by the accumulated weights $\alpha_t$ might not always be accurate,
as shown in the dashed box in Fig.~\ref{fig:aif},
AIF shows flexibility in tackling this issue by taking the first frame as the left boundary when extracting the $\bm{c}_j$.

\subsection{Streaming Joint Decoding}
\label{stream-decode}
Due to the proposed AIF mechanism, the LS-Transducer is naturally equipped with streaming decoding. In addition, considering the LS-Transducer uses the CTC branch to help model training, a streaming joint decoding method is proposed which computes an online CTC prefix score synchronously with the LS-Transducer predictions to refine the search space and eliminate irrelevant alignments. 

The standard CTC prefix score is inapplicable in streaming scenarios because it requires a complete speech utterance.
Suppose $g$ is a partial hypothesis, $q$ is a token appended to $g$, and $h=g\cdot q$ is a new hypothesis. When $q$ is a normal vocabulary token (i.e. not end-of-sentence $[{\rm eos}]$),
the CTC prefix score $S_{\rm ctc}$ of $h$ in \cite{8068205} is calculated as:
\begin{eqnarray}
    &p_{\rm ctc}(h,\cdots|\textbf{E})=\sum_{\nu\in(\mathcal{\rm U}\cup[{\rm eos}])}p_{\rm ctc}(h\cdot\nu|\textbf{E})\\
    &S_{\rm ctc}(h, \textbf{E})={\rm log}(p_{\rm ctc}(h,\cdots|\textbf{E}))
\end{eqnarray}
where $p_{\rm ctc}$ refers to the sequence probability\footnote{
See \cite{8068205} for detailed computation of the CTC-based sequence probability.} given by CTC, e.g. $p_{\rm ctc}(h\cdot\nu|\textbf{E})$ represents the probability of the sequence $h\cdot\nu$ given the entire encoder output $\textbf{E}$. In this context, $\nu$ represents all possible non-empty tokens (with ${\rm U}$ denoting normal tokens), and $h\cdot\nu$ means appending $\nu$ to $h$. Therefore, the CTC prefix score is calculated as the accumulated probability of all sequences with $h$ as their prefix \cite{8068205}. 
However, if $q$ (i.e. the last token of $h$) is $[{\rm eos}]$, the CTC score is computed as:
\begin{equation}
S_{\rm ctc}(h, \textbf{E})={\rm log}(\gamma_T^{(n)}(g)+\gamma_T^{(b)}(g))
\end{equation}
where $\gamma_T^{(n)}(g)$ and $\gamma_T^{(b)}(g)$ are the forward probabilities \cite{8068205} of the partial hypothesis $g$ over $T$ frames, with CTC paths ending with a non-blank or blank label, respectively. These CTC prefix scores are computed based on whole encoder output $\textbf{E}$ with $T$ frames, hampering streaming decoding.

To compute an online CTC prefix score for streaming joint decoding, inspired by \cite{9383517}, this paper uses 
$S_{\rm ctc}(h, \textbf{E}_{1:T_h})$ to approximate $S_{\rm ctc}(h, \textbf{E})$, where $T_h$ is the maximum number of encoder output frames accessible when predicting the new hypothesis $h$, which is decided by the accumulated weights $\alpha_t$ of AIF. Therefore, the online CTC prefix score and LS-Transducer prediction are strictly synchronised.

However, since CTC has too much flexibility when learning alignments,
there is no theoretical guarantee that
the CTC spikes and AIF-located boundaries will be synchronised.
When the corresponding CTC spike for token $q$ doesn't appear during
$\textbf{E}_{1:T_h}$, preliminary experiments showed that the online CTC prefix score $S_{\rm ctc}(h, \textbf{E}_{1:T_h})$ would be very likely to predict $[{\rm eos}]$ because $h$ would be considered complete for $\textbf{E}_{1:T_h}$, which could greatly degrade the performance.
Previous work tackles this problem by waiting until the corresponding CTC spike appeared before starting decoding \cite{Miao2019OnlineHC} or switching to decoding the next block of speech if the $[{\rm eos}]$ label is predicted \cite{9383517}. However, these methods are not feasible for the proposed LS-Transducer that needs to compute online CTC scores synchronously.

To solve this issue, a proposed streaming joint decoding method modifies the computation of online CTC prefix scores for $[{\rm eos}]$, which is shown as follows where $h$=$g\cdot [{\rm eos}]$: 
\begin{equation}
S_{\rm ctc}(h, \textbf{E}_{1:T_h})=
\begin{cases}
{\rm log}(p_{\rm ctc}(h,\cdots|\textbf{E}_{1:T_h})),& \text{ $ T_h < T$ } \\
{\rm log}(\gamma_{T_h}^{(n)}(g)+\gamma_{T_h}^{(b)}(g)),& \text{ $ T_h = T$} \label{tlce2}
\end{cases}
\end{equation}
This means that if the speech has not been completely loaded (i.e. $T_h < T$), $h$ will not be considered complete, leading to an extremely low score for $[{\rm eos}]$ because CTC training never encounters the $[{\rm eos}]$ label.
This makes sense because the online CTC prefix score should not consider ending prediction before reading the whole speech utterance.

\begin{algorithm}[t]
\caption{Modified Online CTC Prefix Score} 
\footnotesize 
\begin{algorithmic}[1]
\Require {$h, \textbf{E}_{1:T_h}$}
\Ensure {$S_{\rm ctc}$}
\State $g, q \gets h$: Split $h$ into the last label $q$ and the rest $g$
\If {$q=[{\rm eos}]$ \textbf{and} $T_h = T$}
    \State \Return ${\rm log}(\gamma_{T_h}^{(n)}(g)+\gamma_{T_h}^{(b)}(g))$
\Else
    \State$
    \gamma_{1}^{(n)}(h)\gets
\begin{cases}
    p(z_1=q|\textbf{E}_{1:T_h})),& \text{if $g=[{\rm sos}]$ } \\
0,& \text{otherwise}
\end{cases}
    $
    \State $\gamma_{1}^{(b)}(h)\gets 0$
    \State $\Psi \gets \gamma_{1}^{(n)}(h)$
    \For {$t=2\cdot\cdot\cdot T_h$}
        \State$
    \Phi \gets \gamma_{t-1}^{(b)}(g) + 
\begin{cases}
    0,& \text{if last($g$)$=q$ } \\
\gamma_{t-1}^{(n)}(g),& \text{otherwise}
\end{cases}
    $
        \State $\gamma_{t}^{(n)}(h) \gets (\gamma_{t-1}^{(n)}(h)+\Phi)p(z_t=q|\textbf{E}_{1:T_h})$
        \State $\gamma_{t}^{(b)}(h) \gets (\gamma_{t-1}^{(b)}(h)+\gamma_{t-1}^{(n)}(h))p(z_t={\rm blank}|\textbf{E}_{1:T_h})$
        \State $\Psi \gets \Psi + \Phi\cdot p(z_t=q|\textbf{E}_{1:T_h})$
    \EndFor
    \State \Return ${\rm log}(\Psi)$
\EndIf
\end{algorithmic}
\end{algorithm}

The detailed procedure for the modified online CTC prefix score is shown in Algorithm 1, which modifies the condition in line 2 compared to the standard CTC prefix score \cite{8068205}. $z_t$ and $p(z_t=q|\textbf{E}_{1:T_h})$ are 
the label and probability for the $t$-th frame.
$[{\rm sos}]$ is start-of-sentence. Other details follow \cite{8068205}.

During streaming joint decoding, score $S_{\rm lst}$ assigned by the LS-Transducer is computed synchronously with $S_{\rm ctc}(h, \textbf{E}_{1:T_h})$
and follows the chain rule:
\begin{equation}
    S_{\rm lst}(h, \textbf{E}_{1:T_h})=\sum_{i=1}^n{\rm log}(p_{\rm lst}(h_i|h_1, \cdots, h_{i-1}, \textbf{E}_{1:T_i}))
\end{equation}
where $p_{\rm lst}$ denotes the predicted probabilities obtained from the final logits output by the joint network, as shown in Fig.~\ref{ls-t}, $n$ is the length of hypothesis $h=g\cdot q$, and $T_i$ is the corresponding right boundary of the $i$-th label as decided by AIF. 
The overall streaming score $S$ is computed as:
\begin{equation}
    S(h, \textbf{E}_{1:T_h}) = \beta S_{\rm ctc}(h, \textbf{E}_{1:T_h})+(1-\beta)S_{\rm lst}(h, \textbf{E}_{1:T_h}) \label{dec}
\end{equation}
where $\beta$ denotes the weight of online CTC scores.
Hence, the streaming scores of the LS-Transducer $S_{\rm lst}(h, \textbf{E}_{1:T_h})$ and the CTC branch $ S_{\rm ctc}(h, \textbf{E}_{1:T_h})$ are strictly synchronised.

\subsection{Prediction Network Adaptation}
With the prediction network of the LS-Transducer performing as an explicit LM, fine-tuning it on text-only data when encountering a domain shift is straightforward. Therefore,
after ASR training and before decoding on an unseen domain, when target-domain text data set $\mathcal{D}$ available, the fine-tuning objective is:
\begin{equation}
    \mathcal{L}_{\text{finetune}} = -\sum_{\bm{Y}\in\mathcal{D}}\sum_{n=1}^{N} {\rm log}~p_{\text{pred}}(y_n|\bm{Y}_{0:n-1}; \theta_{\text{pred}}) \label{finetune}
\end{equation}
where $\bm{Y}=([sos], y_1, ..., y_N)$ is a text sequence belonging to $\mathcal{D}$ and $y_n$ denotes the $n$-th token. $\theta_{\text{pred}}$ represents the parameters of the prediction network, and $p_{\text{pred}}$ denotes the predicted probabilities of the prediction network, which are obtained by applying the softmax function to its output logits, i.e. linear classifier output as shown in Fig~\ref{ls-t}.

\subsection{Specific Implementation}
In this paper, inspired by \cite{9398531}, a simple method is used to compute the weight $\alpha_t$ in AIF by applying a sigmoid function to
the last ${e_{t,d}}$ of each encoder output frame $\bm{e_t}$:\footnote{LS-Transducer is not limited to this method of generating $\alpha_t$, other methods including convolutional or fully-connected layers could be used.}
\begin{equation}
\alpha_t = {\rm sigmoid}({e_{t,d}})
\end{equation}
where $d$ is the dimension of $\bm{e}_t$. Preliminary experiments show that an auxiliary phone-based quantity loss helps model training, so Eq.~\ref{bpe:qua} is written as $\mathcal{L}_{\rm qua} \text{=} |\sum_{i=1}^T\alpha_i \text{-} L| \text{+} |\sum_{i=1}^Tw_i \text{-} P|$, where $w_t = {\rm sigmoid}({e_{t,d-1}})$ and $P$ 
corresponds to the number of phone units in the utterance.
Correspondingly, when AIF extracts the label-level representations $\bm{c}_j$, 
only the other elements of $\bm{e}_t$ are used, i.e. $\textbf{e}_{t,1:d-2}$, so Eq.~\ref{label} of AIF can be expressed as follows in this specific implementation.
\begin{equation}
    \bm{c_j} = {\rm softmax}(\bm{d}_j^{\rm inter}\cdot{ {{\rm FC}(\textbf{E}_{1:T_j,1:d-2})}^{\top}})\cdot{ {\rm FC}(\textbf{E}_{1:T_j,1:d-2})} 
\end{equation}
where {\rm FC} denotes fully connected layers that map $\bm{e}_{t,1:d-2}$ to the same dimension as the $\bm{d}_j^{\rm inter}$. 

\section{Experimental Setup}
\label{setup}
\subsection{Datasets}
ASR models were trained on the LibriSpeech \cite{7178964} data, a read audiobook corpus, and its dev/test sets (i.e. “test/dev-clean/other”) were used for intra-domain evaluation. The source-domain text data included training set transcripts and LibriSpeech LM training text.
To evaluate the domain adaptation capability of the LS-Transducer, two out-of-domain test corpora were used. 
The first corpus consisted of TED-LIUM2 \cite{rousseau-etal-2014-enhancing} dev/test sets, comprising spontaneous lecture-style data. For target-domain adaptation text,
the training set transcripts and TED-LIUM2 LM training text were used.
The second corpus was AESRC2020 \cite{9413386} dev/test sets, containing human-computer interaction speech commands, and the training set transcriptions were used as the target-domain text data.

The models and experimental evaluations were implemented based on the ESPnet2 \cite{Watanabe2018ESPnet} toolkit.
Raw speech data was used as input, and 1000 ASR modelling units were used as text output, including 997 BPE \cite{gage1994} units and 3 non-verbal symbols (i.e. blank, unknown-character and start/end-of-sentence).

\subsection{ASR Model Description}
\subsubsection{Standard Neural Transducer Models}
Three standard Transformer transducer (T-T) \cite{Zhang2020TransformerTA} models, built with streaming wav2vec 2.0 encoders and different prediction networks, were compared to the proposed LS-Transducer.
The T-T with an
embedding layer as the prediction network is denoted the Stateless-Pred T-T (319M parameters); the T-T with a 6-layer 1024-dimensional LSTM prediction network is referred to as the LSTM-Pred T-T (370M parameters); and the T-T with a 6-layer unidirectional Transformer prediction network (1024 attention dimension, 2048 feed-forward dimension, and 8 heads) is called as the Transformer-Pred T-T (371M parameters). 
Streaming wav2vec 2.0 encoders, based on a ``w2v\_large\_lv\_fsh\_swbd\_cv" \cite{hsu21_interspeech}, were implemented using a chunk-based mask \cite{li20_interspeech} to enable streaming processing, with a 320 ms average latency. 
Inspired by \cite{Zhao2022FastAA, 9688251}, the three standard T-T models also used the CTC branch with 0.3 weight to help training.

\subsubsection{LS-Transducer}
The proposed LS-Transducer (373M parameters) had the same streaming wav2vec2.0 encoder as the three standard T-T baseline models and had a unidirectional Transformer prediction network, which was the same as that of Transformer-Pred T-T. The intermediate output from the $3$rd sub-layer of the prediction network was employed as the query for the AIF mechanism.
The linear layers in Fig.~\ref{ls-t} mapped dimensions from 1024 to 1000.
In Eq.~\ref{obj}, the CTC loss was computed based on $\bm{e}_{t,1:d-2}$ where $d=1024$, and $\gamma$ and $\mu$ were respectively set to 0.5 and 0.05. In Eq.~\ref{dec}, $\beta$  was set to 0.3 except for TED-LIUM2 which was 0.4.

\subsubsection{Offline AED Model}
An offline Transformer-based AED model (394M parameters) was also built to compare with the online LS-Transducer. It used the same streaming wav2vec 2.0 encoder but underwent offline training and was decoded using offline CTC/attention joint decoding \cite{8068205}.

\subsubsection{Related Variants of Standard Neural Transducer}
\label{variant}
Building upon the Transformer-Pred T-T structure,
both factorised T-T \cite{Chen2021FactorizedNT} (372M parameters) and HAT \cite{9053600} (371M parameters) were implemented with the same encoder and prediction network (called the vocabulary predictor in factorised T-T). For the factorised T-T, the embedding layer from the vocabulary predictor was directly used as the blank predictor.

\subsection{LM Model and Text Adaptation}
A source-domain 6-layer Transformer LM was trained on the source-domain text data
for 25 epochs and fine-tuned on the target-domain text for an additional 15 epochs as a target-domain LM. 
The trained source-domain LM was used to initialise the prediction network of the LS-Transducer but not for the three standard T-T models because this didn't help enhance performance \cite{9054419}. ASR training was for 40 epochs.
When adapting the LS-Transducer prediction network as in Eq.~\ref{finetune}, the first 3 layers were fixed, and the rest were fine-tuned on the target-domain text data with 50 epochs for AESRC2020 and 20 epochs for TED-LIUM2 data.
Shallow fusion \cite{chorowski2015attention} was used with a 0.2 LM weight if using the target-domain LM for domain adaptation. A beam size of 10 was used during inference. 

\section{Experimental Results}
\label{results}
The LS-Transducer was compared to the standard T-T models for both intra and cross-domain scenarios. Ablation studies were conducted to evaluate the effectiveness of AIF, streaming joint decoding, and prediction network initialisation. Several related methods were also implemented and experimentally compared to the LS-Transducer. 

\begin{table}[t]
    \caption{Initial experiments: intra-domain WER on LibriSpeech dev/test sets for online transducer models trained from LibriSpeech-100h (train-clean-100 set).}
  \label{tab:ls100-nt}
  \centering
  \setlength{\tabcolsep}{0.5mm}
  \renewcommand\arraystretch{1.35}
  \begin{tabular}{l | c |c| c| c}
    \Xhline{3\arrayrulewidth}
     \multirow{2}{*}{Online ASR Models}&\multicolumn{2}{c|}{Test}&\multicolumn{2}{c}{Dev}\\
     \cline{2-5}
     &{clean}&{other}&{clean}&{other} \\
     \hline
     (Offline) W2v2 Transducer \cite{yang2022knowledge}&5.2 &11.8& 5.1&12.2\\
     (Offline) Conformer Transducer \cite{albesano22_interspeech}&5.9& 16.9&--&--\\
     Chunked Conformer Transducer \cite{albesano22_interspeech}&6.8& 20.4&--&--\\
     \hline
     (Offline) AED Model&4.4&11.3&4.2&11.3\\
     \hline
     Stateless-Pred T-T&5.6&12.6&5.5&12.6\\
    LSTM-Pred T-T&5.3&12.5&5.1&12.5\\
     HAT \cite{9053600}&5.4&12.2&5.1&12.1\\
    Factorised T-T \cite{Chen2021FactorizedNT}&5.4&12.4&5.3&12.4\\
    Transformer-Pred T-T&5.1&12.0&4.9&12.0\\
    Proposed LS-Transducer&\textbf{4.4}&\textbf{11.0}&\textbf{4.1}&\textbf{10.8}\\
    \Xhline{3\arrayrulewidth}
  \end{tabular}
  \begin{tablenotes}
  \footnotesize
  \item{\hspace{-4.0mm}*}{Note HAT and Factorised T-T results were generated in this paper.}
  \end{tablenotes}
\end{table}

\subsection{Initial Experiments}
\label{initial}
Initial experiments were conducted on the LibriSpeech-100h set and the  ASR results are listed in Table~\ref{tab:ls100-nt}, in which our models showed good results on the LibriSpeech-100h benchmark compared to various recent work. In addition, among the three standard T-T models, the Transformer-Pred T-T performed the best,
showing that a strong Transformer-structured prediction network is very helpful for the neural transducer to achieve high ASR accuracy.
Moreover, compared to the strong Transformer-Pred T-T, HAT \cite{9053600} and factorised T-T \cite{Chen2021FactorizedNT} slightly degraded the intra-domain performance. 
However, the proposed LS-Transducer still clearly surpassed the strong Transformer-Pred T-T model with up to 16.3\% relative WER reduction (WERR). Moreover, the online LS-Transducer even 
slightly outperformed
the offline AED model, showing the advantages of the LS-Transducer, including that the prediction network performs as an explicit LM so that it can be easily initialised by a trained source-domain LM. This initialisation technique proved to be highly effective in improving ASR performance for some non-autoregressive E2E models \cite{9398531}, but remains a challenge for auto-regressive E2E models such as Transformer-based AED \cite{Deng2022ImprovingCS} and standard neural transducer \cite{9054419} due to the lack of an explicit LM component.

Given the strong performance shown by the Transformer-Pred T-T, the main experiments were carried out on the LibriSpeech-960h data focusing on the comparison between the Transformer-Pred T-T and LS-Transducer.

\begin{table}[t]
    \caption{Intra-domain WER on LibriSpeech dev/test sets for online transducer models trained from LibriSpeech-960h.}
  \label{main_intra}
  \centering
  \setlength{\tabcolsep}{0.9mm}
  \renewcommand\arraystretch{1.35}
  \begin{tabular}{l | c |c| c| c}
    \Xhline{3\arrayrulewidth}
     \multirow{2}{*}{Online ASR Models}&\multicolumn{2}{c|}{Test}&\multicolumn{2}{c}{Dev}\\
     \cline{2-5}
     &{clean}&{other}&{clean}&{other} \\
    \hline
    ConvT-T \cite{huang20k_interspeech} &3.5&8.3&--&--\\
    Dynamic Encoder Transducer\cite{shi21b_interspeech}&3.5&9.0&--&--\\
    Parallel Encoder Transducer\cite{sudo23c_interspeech}&3.7&9.3&3.5&9.2\\
    \hline
    Transformer-Pred T-T&3.2&8.0&3.0&7.8\\
    \quad+LM Shallow Fusion&3.1&7.7&2.9&7.5\\
    Proposed LS-Transducer&3.0&7.2&2.9&7.4\\
    \quad+LM Shallow Fusion& \textbf{2.7} & \textbf{6.8} & \textbf{2.6} & \textbf{6.7} \\
    \Xhline{3\arrayrulewidth}
  \end{tabular}
\end{table}
\begin{table}[t]
\caption{Cross-domain WER results on TED-LIUM 2 (Ted2) and AESRC2020 (AESRC) for online transducer models trained from LibriSpeech-960h (LS960). SF denoted shallow fusion\cite{chorowski2015attention}.}
  \label{tab:ls100cross-nt}
  \centering
  \setlength{\tabcolsep}{0.6mm}
  \renewcommand\arraystretch{1.35}
  \begin{tabular}{l | c c| c c}
    \Xhline{3\arrayrulewidth}
     \multirow{2}{*}{Online ASR Models}&\multicolumn{2}{c|}{LS960$\Rightarrow$Ted2}&\multicolumn{2}{c}{LS960$\Rightarrow$AESRC}\\
     &{~\,}{Test}&{Dev}&{~\;}{Dev}&{Test} \\
    \hline
    Transformer-Pred T-T&{~\,}12.7&13.1&{~\;}19.0&18.7\\
    \quad+Target-domain LM SF&{~\,}11.9&12.2&{~\;}16.7&16.2\\
    \cline{1-1}
    Proposed LS-Transducer&{~\,}11.7&12.0&{~\;}18.2&17.8\\
    +Adapting Prediction Net&{~\,}10.0&10.3&{~\;}14.9&14.1\\
    \quad++Target-domain LM SF&{~\,}\textbf{9.1}&\textbf{9.6}&{~\;}\textbf{13.6}&\textbf{12.6}\\
    \Xhline{3\arrayrulewidth}
  \end{tabular}
\end{table}

\subsection{Main Experiments}
Table~\ref{main_intra} lists intra-domain ASR results, with E2E ASR models trained on the LibriSpeech-960h corpus, 
our models yielded competitive results on the LibriSpeech-960h benchmark compared to recent work.
The LS-Transducer still outperformed the Transformer-Pred T-T in the high-resource LibriSpeech-960h scenario, with 10\% relative WERR. In addition, when the external source-domain LM was used for the E2E ASR via shallow fusion \cite{chorowski2015attention}, the performance of both models was further improved, where the LS-Transducer still gave a 12.9\% relative WERR compared to the Transformer-Pred T-T.


The TED-LIUM 2 and AESRC2020 dev/test sets were used to evaluate the cross-domain performance of the 
ASR models trained on the LibriSpeech-960h data.
As shown in Table~\ref{tab:ls100cross-nt}, the proposed LS-Transducer gave the best results on both cross-domain corpora, showing that LS-Transducer generalises well rather than overfitting to the source domain.
With the prediction network adapted/fine-tuned using the target-domain text data, further improvements could be gained and surpass the Transformer-Pred T-T model with 21.4\% and 24.6\% relative WERR on TED-LIUM 2 and AESRC2020, respectively. 
Even when shallow fusion \cite{chorowski2015attention} was used for the Transformer-Pred T-T model to improve the cross-domain performance by incorporating an external target-domain LM,
a performance gap of at least 10.8\% relative WERR still existed compared to the LS-Transducer with adapted prediction network. In addition, the LS-Transducer could also use shallow fusion with the external target-domain LM to further boost cross-domain accuracy.

In summary, the proposed LS-Transducer not only outperforms the standard T-T models within the source domain but also exhibits greatly improved domain adaptation capabilities.
This is primarily because the prediction network of the LS-Transducer works as an explicit LM, which brings advantages in utilising text-only data.

\subsection{Ablation Studies on AIF}
\begin{table}[t]
    \caption{Ablation studies on the label-level encoder representation generation mechanism: intra-domain WER for LS-Transducer trained on LibriSpeech-100h or LibriSpeech-960h with AIF or normal CIF \cite{9054250}.}
  \label{ablation_aif}
  \centering
  \setlength{\tabcolsep}{0.45mm}
  \renewcommand\arraystretch{1.3}
  \begin{tabular}{ c l  c c c c}
    \Xhline{3\arrayrulewidth}
     \multicolumn{2}{l}{\multirow{2}{*}{Online ASR Models}}&\multicolumn{2}{c}{Test}&\multicolumn{2}{c}{Dev}\\
     \cline{3-6}
     &&{clean}&{other}&{clean}&{other} \\
    \hline
    \multicolumn{2}{l}{\emph{LibriSpeech 100h}} & & & & \\
    &{\ }Transformer-Pred T-T&5.1&12.0&4.9&12.0\\
    &{\ }Proposed LS-Transducer w/ AIF&\textbf{4.4}&\textbf{11.0}&\textbf{4.1}&\textbf{10.8}\\
    &{\ }Proposed LS-Transducer w/ CIF&7.0&13.3&6.5&13.2\\
    \hline
    \multicolumn{2}{l}{\emph{LibriSpeech 960h}} & & & & \\
    &{\ }Transformer-Pred T-T&3.2&8.0&3.0&7.8\\
    &{\ }Proposed LS-Transducer w/ AIF&\textbf{3.0}&\textbf{7.2}&\textbf{2.9}&\textbf{7.4}\\
    &{\ }Proposed LS-Transducer w/ CIF&4.6&9.0&4.3&8.7\\
    \Xhline{3\arrayrulewidth}
  \end{tabular}
\end{table}
Ablation studies were conducted to evaluate the effectiveness of the proposed AIF mechanism.
As shown in Table~\ref{ablation_aif}, when ASR models were trained on the LibriSpeech-100h data, using CIF with LS-Transducer resulted in noticeably inferior performance compared to the strong Transformer-Pred T-T model, which is consistent with the conclusion about CIF in \cite{9688157}. 
The proposed AIF gave much lower WER than CIF \cite{9054250} and played an essential role in enabling the LS-Transducer to surpass the strong Transformer-Pred T-T model. This is consistent with the comparison in Sec.~\ref{aif} that the proposed AIF has several advantages that improve performance over CIF, including no mismatch between training and decoding and enhanced robustness to inaccurate unit boundaries. 

When ASR models were trained on the LibriSpeech-960h corpus, as shown in Table~\ref{ablation_aif},
the LS-Transducer with CIF achieved obvious progress compared to when it was only trained on LibriSpeech-100h data. 
However, it still failed to yield competitive performance compared to the Transformer-Pred T-T. 
Consistent with the LibriSpeech-100h scenario, the LS-Transducer with AIF gave relative WERR between 14.9\% to 34.7\% compared to using CIF, thereby allowing the LS-Transducer to exceed the Transformer-Pred T-T.

\subsection{Ablation Studies on Streaming Joint Decoding}
Ablation studies were also conducted to evaluate the proposed streaming joint decoding method. As shown in Table~\ref{ablation_decode}, the LS-Transducer gave competitive results compared to the strong Transformer-Pred T-T model even without the streaming joint decoding. Moreover, streaming joint decoding could further yield up to 23.1\% relative WERR for the LS-Transducer. This is because the online CTC prefix score can help refine the search space and eliminate irrelevant alignments.
However, when the modification of the online CTC prefix score for $[{\rm eos}]$
proposed in this paper, as in Eq.~\ref{tlce2} or line 2 of Algorithm 1, was not used, as mentioned in Section~\ref{stream-decode}, the performance was greatly degraded. Therefore, the proposed streaming joint decoding method is simple and effective and can ensure strict synchronisation of the online CTC prefix score and the LS-Transducer predictions.

\begin{table}[t]
    \caption{Ablation studies: intra-domain WER for LS-Transducer trained on LibriSpeech-960h with or without the streaming joint decoding.}
  \label{ablation_decode}
  \centering
  \setlength{\tabcolsep}{1.2mm}
  \renewcommand\arraystretch{1.3}
  \begin{tabular}{l  c c c c }
    \Xhline{3\arrayrulewidth}
     \multirow{2}{*}{Online ASR on LS960}&\multicolumn{2}{c}{Test}&\multicolumn{2}{c}{Dev}\\
     \cline{2-5}
     &{clean}&{other}&{clean}&{other}\\
    \hline
    Transformer-Pred T-T&3.2&8.0&3.0&7.8\\
    Proposed LS-Transducer\\
    {\ } w/ streaming joint decoding&\textbf{3.0}&\textbf{7.2}&\textbf{2.9}&\textbf{7.4}\\
    \qquad w/o modification for $[{\rm eos}]$&6.8&9.9&6.0&10.0\\
    {\ } w/o streaming joint decoding&3.9&7.9&3.5&7.8\\
    \Xhline{3\arrayrulewidth}
  \end{tabular}
\end{table}

\begin{table}[t]
    \caption{Intra-domain WER for Transformer-Pred T-T and LS-Transducer trained on LibriSpeech-960h with or without prediction network pre-trained.}
  \label{ablation_pre}
  \centering
  \setlength{\tabcolsep}{0.45mm}
  \renewcommand\arraystretch{1.35}
  \begin{tabular}{ l  c c c c}
    \Xhline{3\arrayrulewidth}
     \multirow{2}{*}{Online ASR on LS960}&\multicolumn{2}{c}{Test}&\multicolumn{2}{c}{Dev}\\
     \cline{2-5}
     &{clean}&{other}&{clean}&{other} \\
    \hline
    Transformer-Pred T-T&3.2&8.0&3.0&7.8\\
    {\ } w/ pre-trained prediction network&4.5&9.6&4.2&9.5\\
    Proposed LS-Transducer&\textbf{3.0}&\textbf{7.2}&\textbf{2.9}&\textbf{7.4}\\
    {\ } w/o pre-trained prediction network&3.5&8.1&3.5&8.0\\
    \Xhline{3\arrayrulewidth}
  \end{tabular}
\end{table}

\subsection{Ablation Studies on Prediction Network Initialisation}
\label{ablation:pred}
In addition, considering a trained source-domain LM was used to initialise the prediction network of the LS-Transducer, ablation studies were conducted to evaluate its effectiveness for both the LS-Transducer and Transformer-Pred T-T. As shown in Table~\ref{ablation_pre}, pre-training the prediction network of the Transformer-Pred T-T cannot improve performance but rather harms it, consistent with the conclusion in \cite{9054419}. In contrast, the prediction network initialisation is highly effective for the LS-Transducer because it performs as an explicit LM. Text-only data is normally easier to collect in large quantities, and the source-domain text data in this paper is much larger than the LibriSpeech-960h transcripts, which is why the prediction network initialisation is still effective for this high-resource LibriSpeech-960h data. Hence, the LS-Transducer provides a natural approach to utilise pre-trained LMs in ASR.

\begin{table}[t]
\caption{WER on intra (LS100) and cross-domain (Ted2 and AESRC) test sets for different models trained from LibriSpeech-100h. For cross-domain scenarios, the internal LM of HAT \cite{9053600} was estimated, the vocabulary predictor of factorised T-T \cite{Chen2021FactorizedNT} was adapted on target-domain text, and shallow fusion was used.}
  \label{tab:compare}
  \centering
  \setlength{\tabcolsep}{1.0mm}
  \renewcommand\arraystretch{1.35}
  \begin{tabular}{l | c c| c| c}
    \Xhline{3\arrayrulewidth}
     \multirow{2}{*}{Online ASR on LS100}&\multicolumn{2}{c|}{LS100 Test}&{Ted2}&{AESRC}\\
     &{clean}&{other}&{Test}&{Test} \\
    \hline
    Transformer-Pred T-T Baseline&5.1&12.0&13.6&23.6\\
    HAT \cite{9053600}&5.4&12.2&13.6&23.0\\
    Factorised T-T \cite{Chen2021FactorizedNT}&5.4&12.4&13.3&22.5\\
    \hline
    Proposed LS-Transducer&\textbf{4.4}&\textbf{11.0}&\textbf{11.1}&\textbf{20.7}\\
    \Xhline{3\arrayrulewidth}
  \end{tabular}
\end{table}
\subsection{Comparison with Related Work}
As a further point of comparison, the cross-domain performance of the factorised T-T \cite{Chen2021FactorizedNT} and HAT \cite{9053600} models were compared to the LS-Transducer. 
The intra-domain and cross-domain results are listed in Table~\ref{tab:compare}, as mentioned in Sec.~\ref{initial}, HAT and factorised T-T performed slightly worse than the strong Transformer-Pred T-T in the intra-domain scenario. Nonetheless, leveraging their strengths in domain adaptation, such as internal LM estimation or adaptation, mitigates this gap and results in improved cross-domain performance compared to the Transformer-Pred T-T.
However, the proposed LS-Transducer still significantly outperformed the HAT and factorised T-T in both intra and cross-domain scenarios with relative WERRs between 8.0\% and 18.5\%.

The WER improvement brought by the LS-Transducer 
over the HAT and factorised T-T is statistically significant at the 0.1\% level according to a matched-pair sentence-segment word error statistical test \cite{115546}.

\section{Conclusions}
\label{conclusions}
This paper proposes a label-synchronous neural transducer (LS-Transducer), which offers a natural solution to domain adaptation for online ASR. The LS-Transducer
does not require the prediction of blank tokens and it is therefore easy to adapt the prediction network on text-only data. An Auto-regressive Integrate-and-Fire (AIF) mechanism is designed to generate a label-level encoder representation before being combined with the prediction network output while still allowing streaming. 
In addition, a streaming joint decoding method is proposed to refine the search space during beam search while maintaining synchronisation with the AIF.
Experiments showed that the proposed LS-Transducer had superior ASR performance and
effective domain adaptation capabilities, exceeding standard neural transducers with
12.9\% and 24.6\% relative WER reductions in intra-domain and cross-domain scenarios respectively.

\ifCLASSOPTIONcaptionsoff
  \newpage
\fi



\bibliographystyle{IEEEtran}
\bibliography{./IEEEexample}
%

%

%







\end{document}